\documentstyle[12pt,fleqn]{article}
\renewcommand{\baselinestretch}{1.2}

\newcommand{\bea}{\begin{eqnarray}}
\newcommand{\beq}{\begin{equation}}
\newcommand{\eea}{\end{eqnarray}}
\newcommand{\eeq}{\end{equation}}
\newcommand{\nnu}{\nonumber}
\newcommand{\di}{\mbox{d}}
\newcommand{\spav}[1]{\parbox{1mm}{\vspace*{#1}}}

\begin{document}

\begin{titlepage}
\begin{flushright}
CERN-TH.6927/93
\end{flushright}
\spav{1.2cm}\\
\begin{center}
{\Large\bf The Rate Difference between the Weak Decays}\\
 {\Large\bf of $t$ and  $\bar{t}$ in Supersymmetry}\\
\spav{1.5cm}\\
{\large Ekaterina Christova}
\spav{1cm}\\
{\em Institute of
Nuclear Research and
 Nuclear Energy}\\
{\em Boul. Tzarigradsko
Chaussee 72, Sofia 1784, Bulgaria.}
\spav{1cm}\\and
\spav{1cm}\\
 {\large Marco Fabbrichesi}
\spav{1cm}\\
{\em CERN, Theory Division}\\
{\em CH-1211 Geneva 23, Switzerland}\\
\spav{1.5cm}\\
{\sc Abstract}
\end{center}
We find that the  $CP$-violating
asymmetry $$ \frac{\Gamma \left(t
\rightarrow bW^+ \right) - \Gamma \left(\bar{t} \rightarrow
\bar{b}W^- \right)}{\Gamma \left(t
\rightarrow bW^+ \right) + \Gamma \left(\bar{t} \rightarrow
\bar{b}W^- \right)} $$ at the one-loop order
within the minimal supersymmetric extension of the standard model
is of the order of few  per cent
for maximal $CP$ violation. It could be measured by
considering the rate difference in the one-lepton events.
 \vfill
\spav{.5cm}\\
CERN-TH.6927/93\\
 July 1993

\end{titlepage}

\newpage
\setcounter{footnote}{0}
\setcounter{page}{1}

{\bf 1.}  The weak decay
of the $t$ quark has
been increasingly advocated as a promising process for testing
violations of $CP$ invariance that could arise in extending the standard
model~\cite{tt}.

At the supercolliders, as well as at the Next Linear Collider (NLC), both
$t$ and $\bar{t}$ will be copiously produced and their decay modes
studied. Because of its large mass~\cite{CDF}, the $t$ decays before
forming any hadronic bound state. Events in which three or more jets are
hard, there is missing transverse energy, and a lepton is identified in
the final states can in principle be used to study the difference in the
rate between the weak decays of the $t$ and  $\bar{t}$ quarks. At the
NLC, a sensitivity of $10^{-4}$ in branching ratios is not
unconceivable~\cite{Kane}.

$CP$ invariance can be
violated in the minimal,
supersymmetric extension of the standard model~\cite{MSSM} to a larger
degree  than  in the standard model.
It is therefore of some interest to estimate the size of the
$CP$-violating asymmetry

\beq
\xi_{CP} \equiv \left| \frac{\Gamma \left(t
\rightarrow bW^+ \right) - \Gamma \left(\bar{t} \rightarrow
\bar{b}W^- \right)}{\Gamma \left(t
\rightarrow bW^+ \right) + \Gamma \left(\bar{t} \rightarrow
\bar{b}W^- \right)} \right| \, ,\label{asymmetry}
\eeq
which can be induced at the one-loop level by supersymmetry.

Notice that the asymmetry (\ref{asymmetry})
within the standard model  implies $CPT$
violation  if we neglect generation mixing,
the effect of which cannot be larger than $10^{-4}$, the square of
the largest off-diagonal element of the quark mixing matrix.

The
corresponding supersymmetry-induced
asymmetry in the decay of the $W$~\cite{W}  is already ruled out by the
present bounds on the supersymmetrical masses~\cite{PDB}.

\spav{1.5cm}\\

{\bf 2.} Let us then  consider the minimal supersymmetric
extension of the standard model.
We neglect generation mixing. Hence, only three
terms in the supersymmetric Lagrangian  can give
rise to $CP$-violating phases  that cannot be rotated
away~\cite{phases}: The superpotential
contains a complex coefficient $\mu$ in the term bilinear
in the Higgs superfields. The soft supersymmetry-breaking operators
introduce two further complex terms, the gaugino masses $\widetilde{m}$
 and the left- and right-handed squark mixing
term.

The possible supersymmetric one-loop diagrams are depicted in Figs.\ 1
and 2.

Since only the imaginary part of the supersymmetric loop amplitude
 enters the asymmetry (\ref{asymmetry}), the main contribution to it
comes from the diagrams in which one of the two on-shell
internal particles is the lightest
supersymmetric particle, that is the lightest neutralino.
 These are the
diagrams of Fig.\ 1.
The contribution of the other diagrams (see Fig.\ 2) is
either strongly suppressed (for the diagram with a charged Higgs,
which is proportional to the mass of the $b$ quark) or closed by the
current experimental bounds on the corresponding supersymmetric
particles~\cite{PDB}. We also neglect
 squark mixing,
which would only make the calculation more involved without changing the
order of magnitude of the final result.

To build the relevant diagrams, we use
 the Lagrangian
 \beq
{\cal L} = L_{\tilde{q}\tilde{q}W} +
L_{\tilde{q}q\chi^+} + L_{W^-\chi^+\chi^0}
+ L_{\tilde{q}q\chi^0}\, , \label{L}
\eeq
where~\cite{MSSM}
\bea
L_{\tilde{q}\tilde{q}W} &=& -\frac{ig}{\sqrt{2}}\: W^-_\alpha \left(
\tilde{b}^*_L
\stackrel{\leftrightarrow}{\partial^\alpha} \tilde{t}_L \right) + H.c.
 \, ,  \\
L_{\tilde{q}q\chi^+} &=& -\frac{g}{2}\: \sum_i \left\{ \bar{t} \left[
U_{i1}
(1+\gamma_5)  - \frac{m_t}{\sqrt{2}m_W \sin \beta}
V_{i2}^*
(1-\gamma_5) \right] \chi_i^+ \, \tilde{b}_L \right. \nnu \\
& & +  \bar{b} \left[
V_{i1}
(1+\gamma_5) - \frac{m_b}{\sqrt{2}m_W \cos \beta}
U^*_{i2}
(1-\gamma_5) \right] \chi_i^{+c} \, \tilde{t}_L  \nnu \\
& & - \, \frac{m_b}{\sqrt{2}m_W \cos \beta}
U_{i2} \, \bar{t}
(1+\gamma_5)  \chi_i^{+} \, \tilde{b}_R \nnu \\
& &  \left. -
\, \frac{m_t}{\sqrt{2}m_W \sin \beta}
V_{i2} \, \bar{b}
(1+\gamma_5)  \chi_i^{+c} \, \tilde{t}_R \right\}  \\
L_{W^-\chi^+\chi^0} & = & \frac{g}{2} \:
W_\alpha \sum_{k,i} \bar{\chi}^0_k \gamma^\alpha \left[ O^L_{ki}
(1-\gamma_5 ) + O^R_{ki} (1 + \gamma_5) \right] \chi^+_i  + H.c.
\, , \\
 L_{\tilde{q}q\chi^0} & = &  \frac{g}{2} \:\sum_{k,f} \bar{q}_f
\left[ f^f_k (1+\gamma_5 ) - \frac{\sqrt{2}m_f}{2m_W B_f}
N^*_{k,5-f} (1 - \gamma_5) \right]   \chi^0_k
\tilde{q}_{fL}  \\
& & \!\!\!\!\! + \, \frac{g}{2} \: \sum_{k,f} \bar{q}_f
\left[ g^f_k (1-\gamma_5 ) - \frac{\sqrt{2}m_f}{2m_W B_f}
N^*_{k,5-f} (1 + \gamma_5) \right]   \chi^0_k
\tilde{q}_{fR}
+ H.c.  \nnu \, , \label{c}
\eea
and
\bea
f^f_k &\equiv&  -\sqrt{2} \left[
T_{3f} N_{k2} - \tan \theta_W
\left( T_{3f} - e_f \right) N_{k1} \right] \, ;
\qquad g^f_k \equiv \sqrt{2} \tan
\theta_W e_f   N^*_{k1}\, , \nnu \\
 O^L_{ki} & \equiv & -\frac{1}{\sqrt{2}}
N_{k4}V^*_{i2} + N_{k2}V^*_{i1}   \, ,\nnu \\
O^R_{ki} & \equiv & \frac{1}{\sqrt{2}} N_{k3}^*U_{i2} +
N_{k2}^*U_{i1} \, , \qquad B_f = \left( \sin \beta , \, \cos \beta
\right)     \, . \eea
In~(\ref{L}) and below, $\chi^0$ and $\chi^+$ are the
four-component spinors of the neutralino and chargino physical fields,
$\chi_i^{+c}$ are the chargino charge-conjugate  states, $t$ and $b$ are
the quark fields  and $\tilde{t}$, $\tilde{b}$ their scalar partners.
The index $f$ stands for the flavor of, respectively, the $t$ and $b$
quark; therefore,  $T_{3f}$, $e_f$ and $m_f$ are, respectively, the
third component of the weak isospin, the charge and the mass of the
corresponding quark. The mass mixing matrices $N$ for the neutralinos,
 $U$ and $V$ for the charginos, contain the $CP$-violating phases of
$\mu$ and $\widetilde{m}$.

\spav{1.5cm}\\

{\bf 3.} By neglecting the effect of the mass of the $b$ quark, we can
write  the
amplitude for the decay as
 \beq
{\cal M} = \frac{g}{2\sqrt{2}} \bar{u} (p') \Biggl[
\gamma_\alpha \left( 1 - \gamma_5 \right) + {\cal A} \gamma_\alpha
\left( 1 - \gamma_5 \right) + {\cal B} P_\alpha \left( 1 + \gamma_5
\right) \Biggr] u (p) \epsilon^\alpha (q) \, ,
\eeq
where the coefficients $\cal A$ and $\cal B$
contain the radiative corrections;
$p$ is the momentum of the decaying $t$ quark, $p'$ of the $b$ and $P
\equiv p+p'$.
 The width is therefore proportional to
\beq
|{\cal M}|^2 = \frac{g^2}{8} p \cdot p' \left[ \left( -2 +
\frac{m_t^2}{m_w^2} \right) \left( 1 + 2\: \mbox{Re} {\cal A}
 \: \right)
+ 2 \:\mbox{Re} {\cal B} \: m_t \left(-1 +  \frac{m_t^2}{m_w^2} \right)
\right] \, . \label{M2}
\eeq

The supersymmetric one-loop contribution, arising from the diagrams in
Fig.\ 1, can thus be  denoted as two sums
\beq
{\cal A} = A_a + A_b + A_c \qquad\mbox{and} \qquad {\cal B} =
B_a + B_b +B_c \, , \label{as}
\eeq
the subscripts following the labelling
of the diagrams in Fig.\ 1.  A
straightforward computation by means of~(\ref{L}) gives
 \bea
A_a & = & \sqrt{2} g^2 V_{i1} \Biggl\{
\widetilde{m}^+_ i\widetilde{m}^0_k O^{L}_{ki} f^{t*}_k I
-  O^{R}_{ki} f^{t*}_k \left[
 m_t^2 (a_1 + a_2 + c_4 +2c_1 +c_2 -4c_3) \Biggr. \right. \nnu \\
 & &  -\left. \Biggl. 2 p \cdot p' (c_4 - c_2) \Biggr]
  -
 \frac{\sqrt{2} m_t^2}{2 \sin \beta} N_{k4}^* \left[
\frac{\widetilde{m}^+_i}{m_W}  O^{L}_{ki} ( a_1 + a_2 + I) -
\frac{\widetilde{m}^0_k}{m_W} O^{R}_{ki} (a_1 + a_2) \right]
 \right\} \nnu  \\
A_b & = & -2g^2 m_t^2 f^b_k f^{t*}_k c_3' \nnu \\
A_c & = &  g^2 \frac{m_t^2}{ \sin \beta} V_{i2} \Biggl\{
g^{*t}_k \left[
\frac{\widetilde{m}^+_i}{m_W} O^{R}_{ki} ( a_1 + a_2 +I) -
\frac{\widetilde{m}^0_k}{m_W} O^{L}_{ki} (a_1 + a_2) \right]
\Biggr.
 \nnu \\
& & + \left.
 \frac{\sqrt{2}}{2 \sin \beta} N^*_{k4}
\left[
 O^{L}_{ki}
\left(
\frac{ m_t^2}{m_W^2} (a_1 + a_2 + c_4 +2c_1 +c_2 -4c_3 )
- 2 \frac{p \cdot p'}{m_W^2} (c_4 -
c_2) \right) \right. \right. \nnu \\
 & & \Bigl. \left.
 -\frac{\widetilde{m}^+_i \widetilde{m}^0_k}{m_W^2} O^{R}_{ki}  I
\right]
 \Biggr\}
\eea
and
\bea
B_a & = & \sqrt{2} g^2 V_{i1} m_t \Biggl\{
 O^{R}_{ki} f^{t*}_k
\left(
 a_1 + a_2 +2c_1 + 2c_2 \right)
 \Biggr.
\nnu \\
& & + \left.\frac{\sqrt{2}}{2\sin \beta}
  N_{k4}^* \left[
 \frac{\widetilde{m}^+_i}{m_W} O^{L}_{ki} ( a_1 - a_2) -
\frac{\widetilde{m}^0_k}{m_W} O^{R}_{ki} (a_1 + a_2)
\right]
\right\} \nnu
\\
 B_b & = & - g^2 m_t f^b_k \left[
 f^{t*}_k ( 2c_1' +2c_2' +a_1' +a_2')
- \frac{\sqrt{2}
\widetilde{m}^0_k}{2m_W \sin \beta} N_{k4}^* (2 a_2' +
I')  \right] \nnu \\
B_c & = &  - g^2 \frac{m_t}{ \sin \beta} V_{i2} \left\{
g^{*t}_k \left[
\frac{\widetilde{m}^0_k}{m_W} O^{L}_{ki} ( a_1 + a_2) -
\frac{\widetilde{m}^+_i}{m_W} O^{R}_{ki} (a_1 - a_2) \right]
\right.
\nnu \\
& & - \left.
 \frac{\sqrt{2}}{2 \sin \beta} \frac{m_t^2}{m_W^2}N^*_{k4}
\left[
 O^{L}_{ki}
(a_1 + a_2  +2c_1 +c_2)
\right]
\right\}
\eea

The coefficients $I$, $a_i$ and $c_i$, as well as the primed
ones, are defined by the  loop
momentum integrals as follows:
\bea
a_1 & = & \frac{A}{q^2} - B \frac{P\cdot q}{\widetilde{P}^2 q^2} \nnu \\
a_2 & = & \frac{B}{\widetilde{P}^2} \, ,  \nnu \\
c_1  &= & \frac{E}{q^2\widetilde{P}^2} - \frac{3}{2} \frac{P\cdot q}
{q^2\widetilde{P}^2}
\left[ \frac{C}{\widetilde{P}^2} - \frac{1}{3} \left( F -\frac{D}{q^2}
\right) \right] \nnu  \\
c_2 & = & \frac{3}{2} \frac{1}{\widetilde{P}^2}
\left[ \frac{C}{\tilde{P}^2} - \frac{1}{3} \left( F -\frac{D}{q^2}
\right) \right]  \nnu \\
c_3 & = & -\frac{1}{2m_t^2} \left[ \frac{C}{\widetilde{P}^2} -  F
+\frac{D}{q^2} \right]  \\
c_4 & = & -\frac{1}{2q^2} \left[ -2\frac{D}{q^2}
+4 \frac{P\cdot q}{q^2 \widetilde{P}^2} E - \frac{C}{\widetilde{P}^2}
\left( 1 + 3\frac{(P\cdot q)^2}{q^2\widetilde{P}^2} \right) \right.
\nnu \\
& & \left. +   \left( 1 + \frac{(P\cdot q)^2}{q^2\widetilde{P}^2}
\right)\left( F - \frac{D}{q^2} \right) \right] \label{b}\nnu \, ,
\eea
where
\bea
\lefteqn{\{I,\,A,\,B,\,C,\,D,\,E,\,F\}  \equiv
}\nnu \\
 & &
8\pi \, \int  \frac{\di ^4 k}{(2\pi)^4} \frac{ \{1,\, (k \cdot q),\,
(k \cdot \tilde{P}),\,(k \cdot \tilde{P})^2, \,
(k \cdot q)^2,\, (k \cdot q)(k \cdot \tilde{P}), \,
k^2 \} }{(k^2
-m^2_1)[(k+p')^2 -m^2_2][(k+p)^2-m^2_3]} \, ,
\eea
each coefficient being defined by the corresponding term inside the
curly brackets.

In the integrals above \beq
m_1 = \widetilde{m}_{\tilde{q}} \qquad m_2 = \widetilde{m}^+_i
\qquad m_3
= \widetilde{m}^0_k \eeq
for the diagrams of Figs.\ 1(a) and 1(c), and
\beq
m_1 = \widetilde{m}^0_k \qquad m_2 = \widetilde{m}_{\tilde{q}}
  \qquad m_3 =
\widetilde{m}_{\tilde{q}}
\eeq
for the diagrams of Fig.\ 1(b), thus giving rise to the primed
coefficients; the two orthogonal momenta are defined as follows:
 \beq
q = p' - p  \qquad \mbox{and} \qquad \widetilde{P} = P -
\frac{P \cdot q}{q^2} q \, .
\eeq

The asymmetry (\ref{asymmetry}) is now readily obtained
by using (\ref{M2}) and is

\beq
\xi_{CP} = 2 \left| \mbox{Re} {\cal A} +
\frac{-1 + m_t^2/m_W^2} {-2 + m_t^2/m_W^2} m_t \:\mbox{Re} {\cal B}
\right| \, . \label{20}
\eeq

Equation (\ref{20}) shows that, in order to have a $\xi_{CP}$ different
from zero, we need, at the same time, a non-vanishing absorptive part of
the loop integrals and a $CP$-violating imaginary coupling in the
Lagrangian.

\spav{1.5cm}\\

{\bf 4.} The contribution of any of the possible supersymmetric
$CP$-violating phases can thus be computed. However, we would like to
have a reliable estimate of the effect without having to commit
ourselves to a definite model of supersymmetry breaking. A possible way
is the following.  We consider
 the case in which there is  one common
 phase $\delta_{CP}$ for both the chargino and the neutralino mixing
matrices.
Accordingly, we can factorize out the phase and take the matrix elements
to be all equal. Because of the threshold in the absorptive part of the
loop integral, in the sum over the neutralino states we keep only the
lightest neutralino; we sum over all chargino states, which we take to
be degenerate in mass (and therefore giving no contribution to
 $\delta_{CP}$).

We still have many terms. To obtain a reliable
 estimate,
we use Table 1, where we
have computed the coefficients $I$, $a_i$, $c_i$ and the primed ones
numerically and listed the  values for a
possible choice of the supersymmetric masses and different values of
$m_t$.

 For maximal
$CP$ violation ($\sin \delta_{CP} = 1$), we take  an averaged value
$|V_{i1,2}O^{L,R} f_k| =  |V_{i2}O^{R,L}
g_k| = |V_{i1,2} N_{k4} O^{L,R}|= 1/4$ and $|N_{k4}f_k| = |f_k f_k| =
1/2$ for the mixing matrix elements.
 For $\sin^2\beta =  1/2$,
 $m_t \simeq 130$ GeV, $m_{\tilde{q}} = m_{\chi^+} =100$
GeV and $m_{\chi^0}=18$ GeV (at the exprimental bound),
 we obtain
\beq
 \xi_{CP} = 1.80 \times \alpha_w \simeq 0.05 \, ,
\eeq
that is, an asymmetry of five per cent.

 The asymmetry does not
vary by much within the present possible
 experimental range of $m_t$; while the two terms
$\cal A$ and $\cal B$ independently grow for larger values of $m_t$,
they enter in the asymmetry  with the opposite sign, so
that, for example at  $m_t =
150$ GeV, the asymmetry  is slightly smaller, being of three per cent.
It becomes smaller and  eventually vanishes as we approach the threshold,
where the masses of the on-shell supersymmetric particles are taken
close to the value of the mass of the $t$ quark.

The study of such an asymmetry can be useful in providing
 indirect evidence for
the existence of supersymmetry and in setting  new bounds on
the size of supersymmetric phases~\cite{bounds}.

 \spav{1.5cm}\\
E.C.'s  work has been partially supported by the Bulgarian National
Science Foundation, Grant Ph-16.

\newpage
\renewcommand{\baselinestretch}{1}

\newpage
\begin{scriptsize}

\begin{table}
\begin{tabular}{|c||c|c|c|c|c|c|c|}
\hline \hline
$m_t$ & $I$ & $a_1$& $a_2$&$c_1$&$c_2$&$c_3$&$c_4$ \\ \hline
120 &$ -2.0 \times 10^{-5}$ & $-8.2 \times 10^{-6}$ &
 $-8.4 \times 10^{-6}$ &
$-3.3 \times 10^{-6}$ & $-3.5 \times 10^{-6}$ & $4.4 \times 10^{-8}$ &
 $1.2 \times 10^{-5}$ \\ \hline
130 & $-4.0 \times 10^{-5}$ & $-1.5 \times 10^{-5}$ &
 $-1.6 \times 10^{-5}$ &
$-4.5 \times 10^{-6}$ & $-6.7 \times 10^{-6}$ & $5.3 \times 10^{-7}$ &
 $3.1 \times 10^{-5}$ \\ \hline
150 & $-4.7 \times 10^{-5}$ & $-1.4 \times 10^{-5}$ &
 $-1.8 \times 10^{-5}$ &
$-8.3 \times 10^{-7}$ & $-7.3 \times 10^{-6}$ & $1.6 \times 10^{-6}$
& $3.6 \times 10^{-5}$ \\
\hline \hline
$m_t$ & $I'$ & $a_1'$& $a_2'$&$c_1'$&$c_2'$&$c_3'$&$c_4'$ \\ \hline
120 &$ 1.6 \times 10^{-5}$ & $1.4 \times 10^{-6}$ &
 $1.3 \times 10^{-6}$ &
$1.3 \times 10^{-7}$ & $1.1 \times 10^{-7}$ &
 $-3.5 \times 10^{-8}$ & $-7.5 \times 10^{-7}$ \\ \hline
130 & $4.5 \times 10^{-5}$ & $6.0 \times 10^{-6}$ & $4.1 \times 10^{-6}$
 &
$1.2 \times 10^{-6}$ & $4.8 \times 10^{-7}$ & $-5.9 \times 10^{-7}$ &
 $-8.9 \times 10^{-6}$ \\ \hline
150 & $1.4 \times 10^{-4}$ & $4.5 \times 10^{-5}$ & $8.6 \times 10^{-6}$
 &
$2.3 \times 10^{-5}$ & $5.9 \times 10^{-7}$ & $-4.8 \times 10^{-6}$ &
 $-1.6 \times 10^{-4}$ \\ \hline
\end{tabular}
\caption{$\widetilde{m}^0 = 18 $ GeV and
 $\widetilde{m}^+ = m_{\tilde{q}}= 100$ GeV}
\end{table}
\end{scriptsize}

\clearpage
\newpage
{\bf Figure captions:}
\spav{1cm}\\
{\bf Fig. 1:} The diagrams relevant in computing $\xi_{CP}$.
\spav{1cm}\\
{\bf Fig 2:} Other possible diagrams, which are closed by threshold;
$\widetilde{\lambda}$ is the gluino, $H$ the charged Higgs.

\end{document}